\documentclass[MSNbibl,nameyear,dvips]{arxstspdf}
\usepackage{flushend}
\usepackage{stfloats}
\usepackage{graphicx}
\volume{26}
\issue{2}
\pubyear{2011}
\firstpage{288}
\lastpage{290}
\doi{10.1214/11-STS349A}
\referstodoi{10.1214/10-STS349}

\makeatletter
\newcommand{\by}{\mathbf{y}}
\newcommand{\btheta}{\bolds{\theta}}

\newtheorem{theorem}{Theorem}
\makeatother

\begin{document}
\begin{frontmatter}

\title{Discussion of ``Estimating Random Effects via Adjustment for Density Maximization'' by C. Morris and R. Tang}
\runtitle{Discussion}
\pdftitle{Discussion of Estimating Random Effects via Adjustment for Density Maximization by C. Morris and R. Tang}

\begin{aug}
\author[a]{\fnms{Claudio} \snm{Fuentes}\ead[label=e1]{cfuentes@stat.ufl.edu}}
and
\author[b]{\fnms{George} \snm{Casella}\corref{}\ead[label=e2]{casella@stat.ufl.edu}}
\runauthor{C. Fuentes and G. Casella}

\affiliation{University of Florida}

\address[a]{Claudio Fuentes is Ph.D. candidate, Department of Statistics, University of Florida, Gainesville, Florida 32611, USA \printead{e1}.}
\address[b]{George Casella is Distinguished Professor, Department of Statistics, University of Florida, Gainesville, Florida 32611, USA \printead{e2}.}

\end{aug}



\end{frontmatter}

We congratulate Morris and Tang for an interesting addition to
empirical Bayes methods, and for tackling a difficult and nagging
problem in variance estimation.  The ADM adjustment appears to bring on
interesting properties, not just in variance estimation but also in
estimation of the means.  In this discussion we want to focus on the
latter topic, and see how the ADM-derived estimators of a normal mean
perform in a decision-theoretic way.  To facilitate this we will stay
with the simple model
\begin{equation}\label{eq:model}
y_i \vert \theta_i \sim N(\theta_i, V),\quad    \theta_i \sim N(0, A).
\end{equation}

\section{The James--Stein Estimator as Generalized Bayes (Not!)}

$\!\!$We first address the comment of Morris and Tang~in Section 2.5, that
the prior $ A \sim \operatorname{Unif}(0, \infty)$ is strong\-ly suggested
because the James--Stein estimator is the posterior mean if we take $A
\sim \operatorname{Unif}(-V, \infty)$.  Professor Morris has noted this before,
and in the interest of understanding, we want to show this calculation
and comment on its relevance.

Writing $\by =(y_1, \ldots, y_k)$ and $\btheta=(\theta_1, \ldots,
\theta_k)$, the posterior expected loss from model (\ref{eq:model}),
with the $A \sim \operatorname{Unif}(-V, \infty)$ prior, is
\begin{eqnarray}\label{eq:bayesrisk}
&&\int_{-V}^\infty \int_{\Re^p} \vert \btheta - \delta(\by)
\vert^2\nonumber\\[-8pt]\\[-8pt]
&&\hphantom{\int_{-V}^\infty \int_{\Re^p}}{}\cdot\frac{e^{-\vert \by -\theta \vert^2/(2V)}}{(2 \pi V)^{k/2}}\frac{e^{-\vert \theta \vert^2/(2A)}}{(2 \pi A)^{k/2}}d\btheta\, dA,\nonumber
\end{eqnarray}
and factoring the exponent in (\ref{eq:bayesrisk}) and writing $B=V/(V+A)$ shows that
\begin{eqnarray*}
\btheta \vert \by, A &\sim& N\bigl((1-B)\by, V(1-B)\bigr),   \\
A \vert \by &\sim&  \biggl(\frac{1}{V+A} \biggr)^{k/2} e^{- ({1}/{(2(V+A))} )\vert \by \vert^2}.
\end{eqnarray*}
The Bayes rules is the posterior mean, which we can calculate as
\begin{eqnarray*}
\mathrm{E} (\btheta \vert \by) &=& \mathrm{E}[\mathrm{E} (\btheta \vert \by, A)]\\
&=& \mathrm{E}[(1-B)\by \vert \by ] = [1-\mathrm{E}(B \vert \by)]\by.
\end{eqnarray*}

We now, very carefully, calculate $\mathrm{E}(B \vert \by)$, yielding
\begin{eqnarray*}
\mathrm{E}(B \vert \by)&\propto& \int_{-V}^\infty  \biggl(\frac{V}{V+A} \biggr) \biggl(\frac{1}{V+A} \biggr)^{k/2}\\
&&\hphantom{\int_{-V}^\infty}{}\cdot e^{- (1/{(2(V+A))} )\vert \by \vert^2}\,dA\\
&=& V \int_{1/V}^\infty t^{k/2-1}e^{-t|\by|^2/2}\,dt\\
&&{} +  V \int_{0}^{1/V} t^{k/2-1}e^{-t|\by|^2/2}\,dt,
\end{eqnarray*}
where we make the transformation $t=1/(V+A)$, with the first integral coming from $A \in (-V,0)$.  Noting that the integrand is the kernel of a chi-squared density, we finally have
\begin{eqnarray}\label{eq:twoparts}
\qquad \mathrm{E}(B \vert \by)&\propto& \frac{V \Gamma(k/2) 2^{k/2}}{(|\by|^2)^{k/2}} [P(\chi^2_{k} \ge |\by|^2/V)\nonumber\\[-8pt]\\[-8pt]
&&\hphantom{\frac{V \Gamma(k/2) 2^{k/2}}{(|\by|^2)^{k/2}} [}
{} +  P(\chi^2_{k} \le |\by|^2/V) ],\nonumber
\end{eqnarray}
where $\chi^2_{k}$ is a chi-squared random variable with $k$ degrees of freedom.  Since the chi-squared probabilities sum to 1, normalizing this expectation (dividing by \hbox{$\frac{\Gamma(k/2-1) 2^{k/2-1}}{(|\by|^2)^{k/2-1}} $}) results in $\mathrm{E}(B \vert \by)= V(k-2)/|\by|^2$, yielding the James--Stein estimator.  There are a number of things to note:
\begin{longlist}[0.]
\item[1.]  If this were a valid calculation, it would contradict such
important papers as Brown (\citeyear{Bro71}) and Strawderman and Cohen (\citeyear{StrCoh71}),
which provided\break  complete characterizations of admissible generalized
Bayes estimators.
\item[2.]  In fact, Strawderman and Cohen [(\citeyear{StrCoh71}),
Section~4.5], explicitly tell us that the James--Stein estimator cannot
be generalized Bayes.
\item[3.]  In fact, the calculation leading to
$\mathrm{E}(B \vert \by)= V(k-2)/|\by|^2$ is invalid.  To see this note
that, starting from (\ref{eq:model}), with $A \sim U(-V, \infty)$, the
prior on $\theta$~is
\[
\int_{-V}^\infty \frac{e^{-\vert \theta \vert^2/(2A)}}{(2 \pi A)^{k/2}} \,dA,
\]
and, even if we take $k$ to be even to avoid complex integration, it is straightforward to verify that the integral over $(-V, 0)$ is infinite.
\end{longlist}

What does this tell us about the James--Stein estimator?  The ``bad''
part of the integral, which leads to the piece in (\ref{eq:twoparts})
corresponding to $P(\chi^2_{k} \ge |\by|^2/V)$, is to be avoided.  We
can informally interpret this as pointing to the region where $
|\by|^2/V$ is small, resulting in shrinkage factors that could be
greater than 1 (in absolute value), and result in the James--Stein
estimator both changing the sign and expanding $\by$. When we lop off
this part, we are led to estimators such as the positive-part
James--Stein estimator, or admissible estimators, like those based on
(32) in Morris and Tang.

\section{Minimaxity of ADM}

The lesson from the previous section is to avoid estimators that do not
control  the shrinker to be between $0$ and $1$.  So we turn  to ADM
and ask if it can do this.  We find, interestingly, that the ADM
approach will, almost automatically, give us a minimax estimator and,
moreover, it controls the shrinker.

$\!$Typically, minimax estimators have been construc\-ted using empirical
Bayes arguments and a bit of customizing, or using formal Bayes
derivations with priors like $A \sim \operatorname{Unif}(0, \infty)$.
The derivation of Morris and Tang in Section 2.7 is a straightforward
differentiation, and we can apply the following theorem.  [This is
Theorem~5.5, Chapter 5, Lehmann and Casella (\citeyear{LehCas98}), and can be traced
back to Baranchik (\citeyear{Bar70}).]

\begin{theorem}
Under model (\ref{eq:model}), the estimator
\[
\delta(\by) =  \biggl(1-\frac{V g(\vert \by \vert)}{\vert \by \vert^2} \biggr) \by
\]
is minimax under the loss $ \vert \btheta - \delta(\by) \vert^2$ if
\begin{longlist}[0.]
\item[1.]  the function $g(\vert \by \vert)$ is nondecreasing,
\item[2.]  $0 \le g(\vert \by \vert) \le 2(k-2)$.
\end{longlist}
\end{theorem}

In the notation of Morris and Tang, we are considering the case $r=0$, and writing $T=\vert \by \vert^2/(2V)$, the ADM shrinkage factor is
\begin{equation}\label{eq:adm}
\quad \hat B = \frac{1}{T} \biggl[\frac{2(m-c+1)T}{T + m+1 +\sqrt{(T-m-1)^2 +4cT}}  \biggr].\hspace*{-10pt}
\end{equation}
Morris and Tang note that $\hat B$ is monotone decreasing in~$T$, but
for minimaxity we need the function in square brackets, which
corresponds to $g(\cdot)$ of the theorem, to be nondecreasing.
\textcolor{black}{ As $T \rightarrow \infty$, the function converges to
$m-c+1$.  If this is the maximum,  and is less than $2(k-2)$, then the
estimator will be minimax.  In fact, it is straightforward (but
tedious) to show that the derivative of the function in square brackets
is always nonnegative, so the function is nondecreasing and the
estimator is minimax. For $c=1$ the bound can be satisfied by taking
$m=(k-2)/2$, for $k \ge 3$.}

\begin{figure*}

\includegraphics{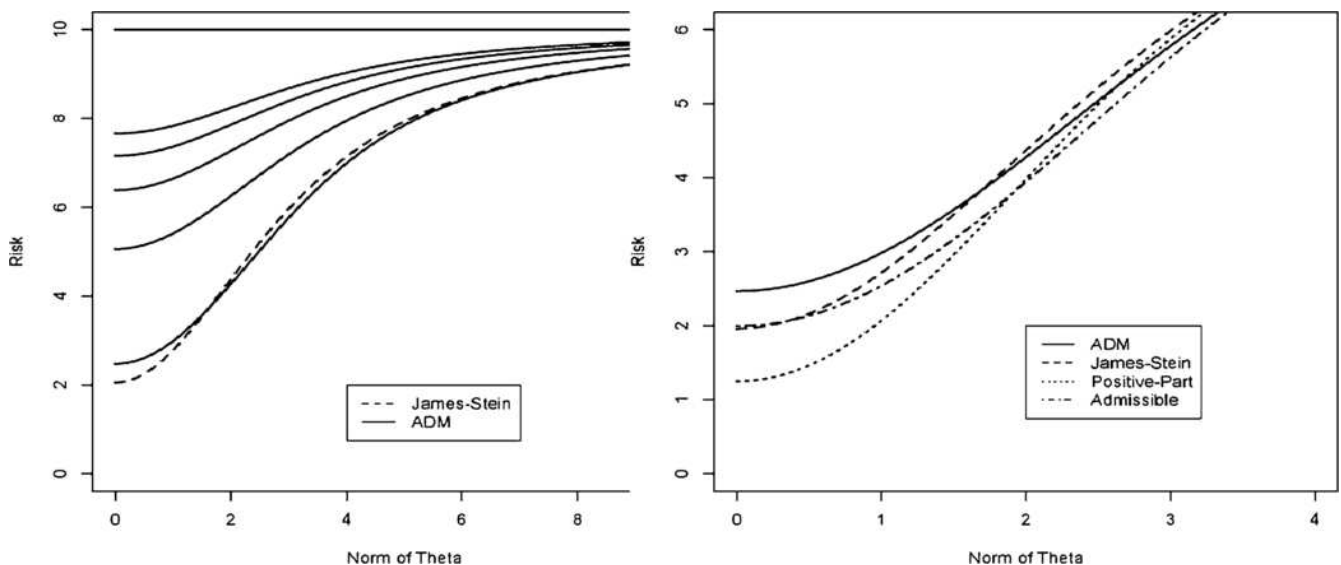}

\caption{For dimension $k=10$, the left panel shows
the risk of the James--Stein estimator (dashed line) and five ADM
estimators (solid lines), for $m=(k-2)/2,(k-2)/4, (k-2)/6, (k-2)/8,
(k-2)/10$.  The risk function increases uniformly as the denominator
increases, so $m^\ast=(k-2)/2$ gives the smallest risk.  The right
panel shows the risk of the ADM estimator with $m^\ast=(k-2)/2$
(solid), the James--Stein estimator (dashed), the positive-part
estimator (dotted), and the admissible estimator with $B$ of (32)
(dash-dot).}\label{fig:ADM}
\end{figure*}

Unfortunately, this is as far as we can go.  The estimator based on
(\ref{eq:adm}), which is reminiscent of a ridge regression estimator,
cannot be admissible.  Again we can trace this back to Strawderman and
Cohen (\citeyear{StrCoh71}), and also Berger and Srinivasan (\citeyear{BerSri78}).  The problem is
that (being a bit informal here)  admissible estimators must be
analytic in the complex plane which is not the case with those based on
(\ref{eq:adm}).

Lastly, we wanted to see the risk performance of ADM.  Morris and Tang
set $m=(k-2)/2$, but there is actually a range of values of $m$ for
which the estimator is minimax.  To clarify, denote $(k-2)/2=m^\ast$,
the Morris and Tang choice, and consider the $m$ in~(\ref{eq:adm}) to
be a variable.  Then, for $c=1$ the estimator is minimax for all $m \le
2(k-2)$.  In Figure~\ref{fig:ADM} we see, in the left panel, the risk
of five ADM estimators, along with the risk of the James--Stein
estimator for comparison.  There we see that the choice of $m$
completely orders the ADM risk, with $m^\ast=(k-2)/2$ being the best
choice, resulting in an estimator with risk similar to that of
James--Stein.  In the right panel we compare the ADM estimator, with
$m^\ast=(k-2)/2$, to the James--Stein estimator, its positive-part
version, and the admissible estimator with $B$ given in (32).  There we
see that ADM compares favorably with the James--Stein estimator, is
uniformly dominated in risk by the admissible estimator, but not by the
positive-part estimator, whose risk crosses that of ADM for large
$|\theta|$.

\section{Is ADM Automatic?}
The automatic appearance of the ADM minimax estimator gives support to
the claim of Morris and Tang that ``ADM maintains the spirit of MLE
while making small sample improvements.'' In fact, examination of the
ADM shrinker $B$, and its risk functions, shows that ``automatic'' ADM
produces an estimator that does not shrink as strongly as either the
admissible estimator or the positive-part and, hence, can have smaller
risk for larger values of the norm of  $\theta$.  It is not clear to
us that such small sample properties as minimaxity will continue to
hold for other models, for example for the estimation of a Poisson
mean, where many similar minimaxity results hold.  However, the results
of Morris and Tang are encouraging and certainly deserve further
investigation.

\section*{Acknowledgment}
$\!$This work is supported by NSF Grant MMS 1028329.


\end{document}